\documentclass[aps,pre,twocolumn,superscriptaddress]{revtex4}

\usepackage[utf8x]{inputenc}
\usepackage{ae} 
\usepackage[T1]{fontenc}
\usepackage[english]{babel}
\usepackage{amsmath}
\usepackage{amsfonts}
\usepackage{amssymb}
\usepackage{array}
\usepackage{graphics}
\usepackage{graphicx}
\usepackage{epsfig}
\usepackage{latexsym}
\usepackage{textcomp}
\usepackage{gensymb}
\usepackage{verbatim}
\usepackage{subfigure}
\usepackage{color}
\usepackage[retainorgcmds]{IEEEtrantools}
\newcommand{\rd}{\mathrm{d}}
\newcommand{\nl}{\nonumber \\}
\newcommand{\mc}{\IEEEeqnarraymulticol}
\newcommand{\la}{\langle}
\newcommand{\ra}{\rangle}
\newcommand{\eps}{\varepsilon}

\DeclareMathOperator{\erfc}{erfc}
\DeclareMathOperator{\sgn}{sgn}

\begin{document}

\title{Critical points of a non-Gaussian random field}
\date{}

\author{T.H. Beuman}
  \affiliation{Instituut-Lorentz for Theoretical Physics, Leiden University, NL 2333 CA Leiden, The Netherlands}
\author{A.M. Turner}
  \affiliation{Institute for Theoretical Physics, Universiteit van Amsterdam, NL 1090 GL Amsterdam, The Netherlands}
\author{V. Vitelli}
  \email{vitelli@lorentz.leidenuniv.nl}
  \affiliation{Instituut-Lorentz for Theoretical Physics, Leiden University, NL 2333 CA Leiden, The Netherlands}

\date{\today}

\begin{abstract}

Random fields in nature often have, to a good approximation, Gaussian characteristics. We present the mathematical framework for a new and simple method for investigating the non-Gaussian contributions, based on counting the maxima and minima of a scalar field. We consider a random surface, whose height is given by a nonlinear function of a Gaussian field. We find that, as a result of the non-Gaussianity, the density of maxima and minima no longer match and calculate the relative imbalance between the two. Our approach allows to detect and quantify non-Gaussianities present in any random field that can be represented as the height of a smooth two-dimensional surface.

\end{abstract}

\maketitle

A wide range of phenomena feature observables that can be regarded as random fields. The cosmic background radiation \cite{cite_Dodelson} is a famous example, but the height profile of a growing surface \cite{cite_KPZ}, medical images of brain activity \cite{cite_Worsley} and optical speckle patterns \cite{cite_Flossmann, cite_Weinrib} also demonstrate this.

In many cases, the fields can be approximated as Gaussian fields, meaning that they have certain properties which are related to the Gaussian (or normal) distribution. This is for example the case when the observable signal is averaged over a large scale, producing approximately Gaussian statistics on account of the central limit theorem. The stochastic properties of such fields have already been the subject of several studies \cite{cite_Longuet2, cite_Berry, cite_Foltin1, cite_Foltin2, cite_Freund, cite_Weinrib}: the density of maxima and minima for instance reflects the amount of field fluctuations at short distances.

Analytical investigations are often restricted to such Gaussian fields. However, phenomena described by nonlinear laws produce non-Gaussian signals. Since these nonlinear effects are usually quite small, the resulting departures from Gaussianity can be tiny. Nevertheless, these non-Gaussianities can offer a key to understanding the interesting nonlinear processes behind the phenomena in question.


If the non-Gaussianity is generated by microscopic nonlinear processes, then some indicator that is sensitive to short distances would be necessary to observe it. Microscopic dynamics do not involve mixing between different regions, so the originally Gaussian field $H(\vec{r})$ simply transforms in a local way, $H(\vec{r})\rightarrow F(H(\vec{r}))$. Provided that this transformation is nonlinear, the new function will have non-Gaussian statistics.


The standard approach to describing the statistics of a random field is to measure its correlation functions. In the case of a random scalar field $h(x,y)$ with Gaussian statistics, its statistical properties are entirely encoded in its two-point correlation function $\la h(x,y) h(x',y') \ra$ (as a function of the distance between $(x,y)$ and $(x',y')$). The higher-order correlation functions can be factorized into two-point correlation functions, by Wick's theorem. A breakdown in these relationships is evidence that the field is not Gaussian.

For example, to see that the field $F(H)$ (as given above) is non-Gaussian, let us calculate its third-order correlation. Such a correlation would vanish (with respect to the mean) for a Gaussian variable. The third-order correlation function at equal points in space is the skewness $\la (F(H)-\la F(H)\ra)^3\ra$. If $F(H)=H+\eps H^2$, then the skewness is easily found to be $12\eps \la H^2\ra^2$. Hence this field is non-Gaussian.

Another way to measure the non-Gaussianity is to directly measure the skewness. The equal-point correlation function should be most sensitive to non-Gaussianity because $F$ is local. This measurement also determines the value of $\eps$, so it gives some information about the dynamics of the nonlinear evolution.

\begin{figure}
  \centering
  \includegraphics{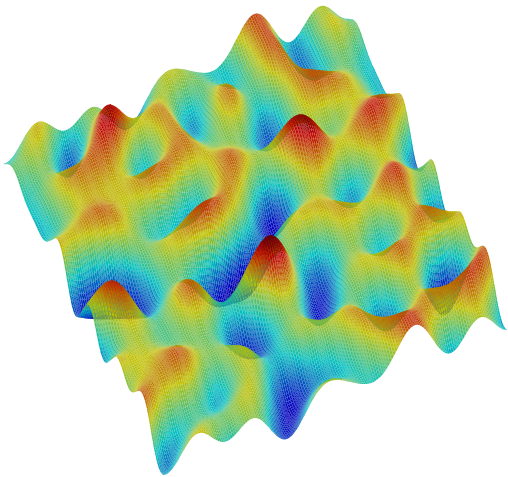}
  \caption{A realization of a Gaussian field with periodic boundary conditions.}
  \label{fig_surface_disk}
\end{figure}

In this paper, we take a geometric approach to tackle this problem. We interpret the scalar field as the height of a surface (see fig.~\ref{fig_surface_disk}) and infer the statistical properties of the signal by studying the stochastic topography of this surface \cite{cite_Kamien}. Such an approach has already been the subject of both theoretical \cite{cite_Dennis1, cite_Longuet2, cite_Berry, cite_Longuet1, cite_Dennis2} and experimental studies \cite{cite_Flossmann}. Here, we focus on the statistical imbalance between peaks and troughs. A test of Gaussianity based on similar ideas has already been applied to the temperature fluctuations in the cosmic microwave background \cite{cite_Heavens, cite_Gupta}. 

We will focus on the \emph{difference} between the densities of maxima and minima. This should also be sensitive to local statistics of the field, but it will be a measurement of the \emph{non-Gaussian} properties in particular, since a Gaussian variable is always symmetric around its mean value. We will study signals of the form $F_{NL}(H)$ where the underlying field $H$ is Gaussian and $F_{NL}$ is any nonlinear function, and we will find that the imbalance can be nonzero, illustrating this approach. Moreover, we show how large the imbalance is \emph{exactly} in relation to the nonlinear perturbation, which allows one to attack the reverse problem: by measuring the difference in density between maxima and minima for a given near-Gaussian field, one can quantify the size of the non-Gaussian component.



The outline of this paper is as follows. In section~\ref{sec_1}, we review the properties of Gaussian fields and introduce the basic notions and notations that we will use. We then demonstrate how the imbalance between maxima and minima can be calculated in section~\ref{sec_2}. Section~\ref{sec_3} is devoted to determining the key ingredient, namely the probability distribution for the values of minima in a Gaussian field. In section~\ref{sec_4} we arrive at the final result, compare it with results from computer generated fields and point out the main features. Finally, section~\ref{sec_5} provides an overview of our findings and their implications.

		\section{Gaussian fields}
	\label{sec_1}

The Gaussian distribution is the archetype of a continuous probability density. It is given by
\begin{equation}
  f(x)  =  \frac1{\sqrt{2\pi}\sigma} \exp\Big( \!-\! \tfrac12 \Big( \frac{x-\mu}{\sigma} \Big)^2 \Big),
\end{equation}
where $\mu$ and $\sigma$ are the expectation value and standard deviation of the stochastic variable respectively. One of its special properties is that the sum of two independent stochastic variables, that adhere to this distribution, is itself also a Gaussian variable, albeit of course with $\mu = \mu_1 + \mu_2$ and $\sigma^2 = \sigma_1^2 + \sigma_2^2$. This property can be considered to be one of the components of the proof of the central limit theorem, which states that -- under some very general conditions -- the sum (or average) of a large number of independent stochastic variables acquires a Gaussian distribution, in the limit that the number goes to infinity \cite{cite_Kampen}. Because of this, many random processes can be well approximated using a Gaussian distribution, e.g.\ the number of times a (fair) coin comes up heads when it is flipped a (large) number of times, or the amount of rain that falls at a certain spot during a year.

A Gaussian random field is an extension of this principle to two dimensions. For instance, one might consider the amount of rain that falls at different places throughout an area rather than a single spot. Upon adding together all the contributions of all rain clouds during the course of a year, one obtains a random field.

Formally, a field is a stochastic function $H(\vec{r})$. The minimum requirement for a Gaussian field is that the probability distribution of $H(\vec{r_0})$ at any point $\vec{r_0}$ has to be described by a Gaussian. More generally, if we consider the values that the field attains at any number of points, $\xi_1 \!=\! H(\vec{r_1}), \, \xi_2 \!=\! H(\vec{r}_2), \, \ldots, \, \xi_n \!=\! H(\vec{r}_n)$, the joint probability distribution has to be of the form
\begin{equation}
  p(\xi_1,\ldots,\xi_n)  \propto  \exp \Big( \!-\! \tfrac12\sum_{i,j}{A_{ij}\xi_i \xi_j} \Big),
  \label{eq_gsn_jpd}
\end{equation}
where $A_{ij}$ are constants. These constants give information about the relative values at different points (which would be useful for example if we wanted to know the distribution of the derivative of the field).

Any well-behaved Gaussian field can be decomposed into Fourier modes, resulting in the sum of an infinite number of wave functions
\begin{equation}
  \psi(\vec{r})  =  \psi_0 + \sum_{\vec{k}} A(\vec{k}) \cos(\vec{k} \cdot \vec{r} + \phi_{\vec{k}}).
  \label{eq_gsn_field}
\end{equation}
This shows how much of the fluctuations occur at each wavelength -- for example, a surface of water might fluctuate with some random waves. If that is due to some external sound at a certain frequency, the Fourier transform will be strongest at the corresponding wavelength.

This procedure may also be turned around -- a Gaussian field may be generated by summing up a large number of Fourier modes. We will now discuss a field that is generated in this way and try to understand how the statistics of the phase factors $\phi_{\vec{k}}$ reflect properties of the field, such as Gaussianity and translational invariance.


The defining characteristic of a Gaussian field is now that the phases $\phi_{\vec{k}}$ are random and completely uncorrelated to each other. Already, by translational invariance, second order correlations between $\phi_k$ and $\phi_{k'}$ are ruled out. If the phases are \emph{completely} independent, then at each individual point $\vec{r}$, $\psi(\vec{r})$ is the sum of an infinite number of independent random numbers between $-1$ and $1$ (as a result of the cosine), each weighted with a factor $A(\vec{k})$. Thus, from the central limit theorem $\psi(\vec{r})$ is a Gaussian random variable. 
In contrast, in a non-Gaussian field the phases \emph{are} correlated, i.e.\ the phases of different modes depend on each other. This mechanism is often called \emph{mode coupling}. 


So far no statements have been made about the function $A(\vec{k})$: it has no influence on the Gaussianity (nor on the homogeneity) of $\psi$. Indeed, this function is a free parameter, called the \emph{amplitude spectrum}. While all Gaussian fields share some general properties, other more specific properties (such as the density of critical points, as we shall see) depend on this amplitude spectrum. For example, when $A(\vec{k})$ is large for vectors $\vec{k}$ with a small norm, the field $\psi$ is dominated by these waves with small wave vectors and hence large wave lengths, resulting in a more slowly varying $\psi$ as compared to a Gaussian field that is dominated by large wave vectors.

There is one more condition that we will pose: next to being homogeneous, we will also only consider fields that are \emph{isotropic}, i.e.\ have rotational symmetry. This is achieved by requiring that $A(\vec{k})$ depends on the magnitude of $\vec{k}$ only, i.e.\ $A(\vec{k}) = A(k)$.

In order to make a clear distinction between Gaussian and non-Gaussian, we will use $H$ to indicate an (isotropic) Gaussian field and $\psi$ for any (homogeneous and isotropic) field. Later we will also use $h$, to indicate a perturbed Gaussian field.

When we have a Gaussian variable $x$ with a certain $\mu$ and $\sigma$, we can make a transformation to $y = \frac{x-\mu}{\sigma}$, which is then a \emph{standard} Gaussian variable, having $\mu = 0$ and $\sigma = 1$. This translation and rescaling has no effect on the overall properties of $x$ and is introduced for convenience. We will apply a similar transformation, by setting $\la H \ra = 0$ and $\la H^2 \ra = 1$. The expectation values are obtained by integrating over all possible values of all random variables, which in this case, are the uniformly distributed phases:
\begin{equation}
  \la \ldots \ra  \equiv  \bigg( \prod_{\vec{k}} \int \! \frac{\rd \phi_{\vec{k}}}{2\pi} \bigg) \ldots
\end{equation}
For our earlier definition eq.~\eqref{eq_gsn_field}, the normalization translates to $H_0 = 0$ and $\sum_{\vec{k}} \tfrac12 A(k)^2 = 1$. This normalization is for the purpose of simplicity only and has no impact on our analysis.

More details on these calculations, as well as additional properties of Gaussian fields and definitions, can be found in appendix~\ref{app_gsn_prop}. There we also demonstrate how the two-point correlation function can be derived from eq.~\eqref{eq_gsn_field}. We also show how the higher-order correlation functions are related to the two-point ones.

Testing whether these relations hold for a given field $\psi$ can reveal whether $\psi$ is Gaussian or not. A more detailed analysis of the correlation functions can provide clues about the nature of the non-Gaussianity.

Although correlation functions provide an excellent approach from a purely mathematical point of view, determining correlation functions for a given realization of a near-Gaussian field $h$ may not always be practical, as it requires precise measurements of $h$ in order to determine the correlation functions with a large enough precision.

In this paper we consider a geometrical test for Gaussianity, which involves counting the number of maxima and minima.

		\section{Maxima versus minima}
	\label{sec_2}

Due to symmetry, a Gaussian field $H$ has as many minima as it has maxima. For a perturbed Gaussian field, like $h = H + \eps H^2$, this may no longer be the case. Therefore the difference in densities of maxima and minima can serve as an indication of non-Gaussianity. We shall now derive what this difference is in the generic case of a field given by $h(\vec{r}) = F_{NL}(H(\vec{r}))$, where $H$ is a Gaussian field and $F_{NL}$ is any (nonlinear) function (e.g.\ the identity plus a perturbation), which depends only on $H(\vec{r})$, i.e.\ the original (unperturbed) value of the field at that same point. This scheme we will refer to as a \emph{local perturbation}.

Transforming the function with $F_{NL}$ does not move maxima and minima around, but it can interchange them, depending on the sign of $F_{NL}' = \rd F_{NL} / \rd H$ at the point in question. To see this, note that maxima and minima, together with saddle points, are \emph{critical points}. The critical points of $h$ are given by
\begin{equation}
  0  =  \vec{\nabla} h(\vec{r})  =  \frac{\rd h}{\rd H} \vec{\nabla} H(\vec{r})  =  F_{NL}'(H) \vec{\nabla} H(\vec{r}).
\end{equation}
We see that the critical points of $H$ and $h$ are the same points; however, the prefactor $F_{NL}'(H)$ may influence the type of critical point. The three types can be distinguished by considering the second derivatives: saddle points have $h_{xx}h_{yy} - h_{xy}^2 < 0$, whereas for maxima and minima (together called \emph{extrema}) this is positive. For maxima, unlike minima, we have $h_{xx} < 0$ -- or $h_{yy} < 0$.

Consider a critical point $\vec{r_0}$ and let $z = H(\vec{r_0})$. The second derivatives of $h$ at $\vec{r_0}$ simply have an extra factor $F_{NL}'(z)$ as compared to the second derivatives of $H$. This has no influence on the sign of $h_{xx}h_{yy} - h_{xy}^2$, therefore, the saddle points (extrema) of $H$ are also saddle points (extrema) of $h$. However, a maximum (minimum) of $H$ is a minimum (maximum) of $h$ when $F_{NL}'(z) < 0$. In order to determine how many extrema will undergo such a transformation, we need to know how often $F_{NL}'(z) < 0$ at such a point.

Let $g(z)$ be the probability density that a certain minimum $\vec{r_0}$ of $H$ has the value $H(\vec{r_0}) = z$. The probability $P$ that a minimum of $H$ becomes
a maximum of $h$ is then 
\begin{equation}
  P  =  \int_{z:F_{NL}'(z)<0} \! \rd z \, g(z).
\end{equation}
For example, if we consider a square perturbation $h = H + \eps H^2$, for which $F_{NL}'(z) = 1 + 2 \eps z$, we have
\begin{equation}
  P  =  \int_{-\infty}^{-\frac1{2\eps}} \! \rd z \, g(z).
\end{equation}
Because of the symmetry of $H$, the maxima are distributed according to $g(-z)$. With that, we can similarly define a probability $Q$ that a maximum becomes a minimum going from $H$ to $h$.

Let $n_0$ be the density of minima (or maxima) of $H$. The density of minima (maxima) of $H$ which are maxima (minima) of $h$ is then $Pn_0$ ($Qn_0$). We can quantify the resulting imbalance in maxima and minima in the dimensionless parameter
\begin{align}
  \Delta n	& \equiv \frac{n_{\mathrm{max}}-n_{\mathrm{min}}}{n_{\mathrm{max}}+n_{\mathrm{min}}}  =  \frac{(1+P-Q)n_0 - (1-P+Q)n_0}{2n_0} \nl
  		& = P-Q  =  \int_{z:F_{NL}'(z)<0} \! \rd z \, \big( g(z) - g(-z) \big).
  \label{eq_maxmin1}
\end{align}
Thus, if we can determine $g(z)$, we can calculate the exact imbalance between the maxima and minima of $h$.

		\section{Distribution of minimum values}
	\label{sec_3}

	\subsection{One dimension}

Let us first consider the probability distribution for minimum values of a Gaussian function on a line. We will then generalize to two dimensions, and afterward, discuss how these distributions depend on the power spectrum. We start with
\begin{equation}
  H(x)  =  \sum_k A(k) \cos(kx + \phi_k).
\end{equation}
The minima are given by $H_x(x_0) = 0$ and $H_{xx}(x_0) > 0$. We would thus like to know the probability density that $H(x_0) = z$, given that $H_x(x_0) = 0$ and $H_{xx}(x_0) > 0$:
\begin{align}
  g(z)	& = p(H(x_\mathrm{min}) = z) \nl
  	& = \frac1{n} \, p(H(x_0) = z \wedge H_x(x_0) = 0 \wedge H_{xx}(x_0) > 0).
  \label{eq_distr1D_concept}
\end{align}
Here $n \equiv p(H_x(x_0) = 0 \wedge H_{xx}(x_0) > 0)$ can be identified as the density of the minima.

We need to determine the joint probability distribution $p(H(x_0), H_x(x_0), H_{xx}(x_0))$ -- since $H$ is homogeneous, $p$ does not depend on $x_0$.

Let us take a closer look at the first derivative
\begin{align}
  H_x(x_0)	& = \sum_k A(k) (-k) \sin(kx_0 + \phi_k) \nl
  		& = \sum_k k A(k) \cos(kx_0 + \phi_k + \tfrac12 \pi).
\end{align}
We see that the expression for $H_x$ still describes a Gaussian: the phases are simply increased by $\tfrac12 \pi$ (modulo $2\pi$) and the spectrum has picked up a factor of $k$. The bottom line is that $H_x(x_0)$ is a Gaussian variable, and it is easy to confirm that the same goes for $H_{xx}(x_0)$ (or any derivative).

We thus have three Gaussian variables. The joint probability distribution of a set of (correlated) Gaussian random variables is given by (compare eq.~\eqref{eq_gsn_jpd})
\begin{equation}
  p(\xi_1,\ldots,\xi_n)  =  \frac1{(2\pi)^{n/2} \sqrt{\det{C}}} \exp \Big( \!-\! \tfrac12\sum_{i,j}{(C^{-1})_{ij}\xi_i \xi_j} \Big).
  \label{eq_gauss_distr}
\end{equation}
Moreover, the coefficients $C$ can be determined measuring the statistics of the field: it is the matrix of correlations
\begin{equation}
  C_{ij}  =  \la \xi_i\xi_j \ra.
\end{equation}

Let us calculate $\la H(x) H_{xx}(x) \ra$ as an example. Again, homogeneity allows us to set $x_0 = 0$ for convenience. We then find
\begin{align}
  & \la H(x_0) H_{xx}(x_0) \ra  =  \la H(0) H_{xx}(0) \ra \nl
  & \qquad = \bigg\la \sum_k A(k) \cos \phi_k \sum_{k'} A(k') (-k'^2) \cos \phi_{k'} \bigg\ra \nl
  & \qquad = \sum_{k k'} A(k)A(k') (-k'^2) \big\la \! \cos \phi_k \cos \phi_{k'} \big\ra \nl
  & \qquad = \sum_{k k'} A(k)A(k') (-k'^2) \tfrac12 \delta_{k k'} \nl
  & \qquad = \sum_k -\tfrac12 A(k)^2 k^2  =  -K_2.
\end{align}
Here we made use of the moment $K_2$ defined in eq.~\eqref{eq_moments}.

An even and an odd derivative of $H$ are always uncorrelated, e.g.
\begin{align}
  & \la H(0) H_x(0) \ra  =  \sum_{k k'} A(k)A(k') (-k') \big\la \! \cos \phi_k \sin \phi_{k'} \big\ra \nl
  & \qquad = \sum_{k k'} A(k)^2 (-k) \big\la \! \cos \phi_k \sin \phi_k \big\ra \delta_{k k'}  =  0.
\end{align}
This is because an even derivative features cosines while an odd derivative has sines, and their product averages to zero, as above.

The final result is that for $H$, $H_x$ and $H_{xx}$ the correlations are
\begin{equation}
  C  =
  \begin{pmatrix}
    1		& 0	& -K_2	\\
    0		& K_2	& 0	\\
    -K_2	& 0	& K_4	\\
  \end{pmatrix}.
\end{equation}
The determinant of $C$ is $K_2(K_4-K_2^2)$ and its inverse is
\begin{equation}
  C^{-1}  =  \frac1{K_2(K_4-K_2^2)}
  \begin{pmatrix}
    K_2K_4	& 0		& K_2^2	\\
    0		& K_4-K_2^2	& 0	\\
    K_2^2	& 0		& K_2	\\
  \end{pmatrix}.
\end{equation}
This gives
\begin{equation}
  \begin{split}
    & p(H, H_x, H_{xx})  =  \frac1{(2\pi)^{3/2}\sqrt{K_2(K_4-K_2^2)}} \\
    & \qquad \times \exp\bigg( \frac{H_x^2}{2K_2} -\frac{K_4H^2+2K_2HH_{xx}+H_{xx}^2}{2(K_4-K_2^2)} \bigg).
  \end{split}
\end{equation}

The plan is now to set $H = z$ and $H_x = 0$ and integrate $p$ over $H_{xx}$. However, one important factor still needs to be added. The probability we have calculated is actually a probability density (since the probability that $H'(x_0)=0$ and $H(x_0)=z$ exactly is zero), and it is not defined with respect to the variables we need. It is defined by fixing a point $x_0$ and determining the probability that $H_x$ vanishes within a certain tolerance at that point:
\begin{equation*}
  \frac{ P(H(x_0) \in [z, z+\rd z] \wedge H_x(x_0) \in [0, \rd H']) }{ \rd z \, \rd H'}
\end{equation*}
Instead, we actually want the probability that there is an \emph{exact} critical point within a certain distance of $x_0$:
\begin{equation*}
  \frac{ P \!
    \begin{pmatrix}
      \exists \: x_m \in [x_0, x_0+\rd x] : \\
      H(x_m) \in [z, z+\rd z] \wedge H_x(x_m) = 0 \\
    \end{pmatrix}
  }{ \rd x \, \rd z }
\end{equation*}
Over the range $\rd x$, $\rd H'$ varies by
\begin{equation}
  \rd H'  =  \bigg| \frac{\partial H_x}{\partial x} \bigg| \rd x  =  |H_{xx}| \rd x.
  \label{eq_Jacobian_1D}
\end{equation}
In order to get the desired probability density with respect to $x$, we need to multiply our current probability density with $|H_{xx}|$.

The probability distribution for the minima is thus given by (see eq.~\eqref{eq_distr1D_concept})
\begin{equation}
  g(z)  =  \frac1{n} \int_0^\infty \! \rd H_{xx} \, p(H = z, H_x = 0, H_{xx}) \, |H_{xx}|.
  \label{eq_distr1D_int}
\end{equation}
The prefactor, featuring the density of minima $n$, can be regarded as a normalization constant and is found by integrating $g(z)$ over the entire $z$-range. This is easily accomplished by taking the expression above and first integrate over $z$, and only then over $H_{xx}$. The result is
\begin{equation}
  \int_{-\infty}^\infty \! \rd z \, g(z)  =  1  \quad \Rightarrow \quad  n  =  \frac1{2\pi}\sqrt{K_4/K_2}.
\end{equation}
The integrand in eq.~\eqref{eq_distr1D_int} is also Gaussian, but it is only integrated over for positive $H_{xx}$, resulting in
\begin{equation}
  \begin{split}
    g(z) =\:	& \sqrt{\frac{1-\lambda}{2\pi}} \exp\bigg( \!-\! \frac1{2(1-\lambda)} \, z^2 \bigg) \\
    		& - \tfrac12\sqrt{\lambda} \, z \, \exp\big( \!-\! \tfrac12z^2\big) \erfc\Bigg( \sqrt{\frac{\lambda}{2(1-\lambda)}} \, z \Bigg).
  \end{split}
  \label{eq_distr1D}
\end{equation}
Here $\erfc$ is the complementary error function
\begin{equation}
  \erfc(x)  \equiv  \frac{2}{\sqrt{\pi}} \int_x^\infty \! \rd t \, e^{-t^2},
\end{equation}
which converges to $1$ as $x$ goes to $-\infty$.
The two parameters $K_2$ and $K_4$ have been merged into a single dimensionless parameter
\begin{equation}
  \lambda  \equiv  \frac{K_2^2}{K_4}  \qquad  (0 \leq \lambda \leq 1).
\end{equation}
Note that we set $K_0 \equiv \la H^2 \ra = 1$ for convenience. In the generic case $K_0 \neq 1$, we have $\lambda = K_2^2/(K_0 K_4)$.
A proof that $\lambda \leq 1$ is derived explicitly in the next section.

	\subsection{Two dimensions}

In two dimensions, the procedure to calculate the distribution of the minima is similar. The minima are defined by the conditions $H_x = H_y = 0$ (defining critical points), $H_{xx}H_{yy} - H_{xy}^2 > 0$ (separating extrema from saddle points) and $H_{xx}, H_{yy} > 0$ (distinguishing minima from maxima). We thus need to find $p(H, H_x, H_y, H_{xx}, H_{yy}, H_{xy})$. This is still a Gaussian joint distribution function.

We start again by determining the correlations, for example (again setting $\vec{r} = 0$ for convenience)
\begin{align}
  \la H_{xx} H_{yy} \ra	& = \sum_{\vec{k}\vec{k'}} A(k)A(k') k_x^2 k'^2_y \big\la \! \cos \phi_{\vec{k}} \cos \phi_{\vec{k'}} \big\ra \nl
  			& = \sum_{\vec{k}\vec{k'}} A(k)A(k') k_x^2 k'^2_y \tfrac12 \delta_{\vec{k}\vec{k'}}  =  \sum_{\vec{k}} \tfrac12 A(k)^2 k_x^2 k_y^2 \nl
  			& = \frac1{2\pi} \int_0^{2\pi} \! \int_0^\infty \! \rd k \rd \theta \, \Pi(k) k^4 \cos^2\theta \sin^2\theta \nl
  			& = \tfrac18 \int_0^\infty \! \rd k \, \Pi(k) k^4  =  \tfrac18 K_4.
\end{align}
In the third line we replaced the sum by an integral and performed it using polar coordinates.

Remember from the one-dimensional case that the correlation of an even and an odd derivative is always zero, because in the calculation we encounter a product of a cosine and a sine, which integrated over the (random) phase yields zero. Based on the calculation method demonstrated above, we can make a more general statement: when the combined number of $x$-derivatives ($y$-derivatives) is odd, the integral over $\theta$ (as above) features a cosine (sine) with an odd exponent; the integral over $\theta$ then gives zero. If we apply this rule to our six variables, we see that $H_x$, $H_y$ and $H_{xy}$ all have no ``compatible match'' in this respect; therefore, they are uncorrelated to all other variables. This allows us to factorize the joint probability distribution
\begin{equation}
  \begin{split}
    & p(H, H_x, H_y, H_{xx}, H_{yy}, H_{xy}) \\
    & \qquad  =  p(H_x) \, p(H_y) \, p(H_{xy}) \, p(H, H_{xx}, H_{yy}).
  \end{split}
\end{equation}
The probability densities of the individual variables are straightforward,
\begin{subequations}
  \begin{align}
    p(H_x)	& = \frac1{\sqrt{\pi K_2}} \exp\Big( \!-\! \frac1{K_2}H_x^2 \Big), \\
    p(H_y)	& = \frac1{\sqrt{\pi K_2}} \exp\Big( \!-\! \frac1{K_2}H_y^2 \Big), \\
    p(H_{xy})	& = \frac2{\sqrt{\pi K_4}} \exp\Big( \!-\! \frac4{K_4}H_{xy}^2 \Big).
  \end{align}
\end{subequations}
For $H$, $H_{xx}$ and $H_{yy}$, we determine the correlation matrix
\begin{equation}
  \renewcommand{\arraystretch}{1.2}
  C  =
  \begin{pmatrix}
    1			& -\tfrac12 K_2	& -\tfrac12 K_2	\\
    -\tfrac12 K_2	& \frac38 K_4	& \frac18 K_4	\\
    -\tfrac12 K_2	& \frac18 K_4	& \frac38 K_4	\\
  \end{pmatrix}.
\end{equation}
The determinant of $C$ is $\frac18K_4(K_4-K_2^2)$ and its inverse is
\begin{equation}
  \renewcommand{\arraystretch}{1.2}
  \begin{split}
    C^{-1}  =\:	& \frac1{K_4(K_4-K_2^2)} \\
    		& \times
    \begin{pmatrix}
      K_4^2	& K_2K_4	& K_2K_4	\\
      K_2K_4	& 3K_4-2K_2^2	& 2K_2^2-K_4	\\
      K_2K_4	& 2K_2^2-K_4	& 3K_4-2K_2^2	\\
    \end{pmatrix}.
  \end{split}
\end{equation}
After some rearranging, eq.~\eqref{eq_gauss_distr} gives
\begin{IEEEeqnarray}{Rll}
  & \mc{2}{l}{ p(H, H_{xx}, H_{yy})  =  \frac1{\pi^{3/2}\sqrt{K_4(K_4-K_2^2)}} } \nl
  & \qquad \times \exp \bigg(	& -\frac{(K_4H+K_2H_{xx}+K_2H_{yy})^2}{2K_4(K_4-K_2^2)} \\
  &				& -\: \frac{(H_{xx}-H_{yy})^2}{2K_4} - \frac{H_{xx}^2 + H_{yy}^2}{K_4} \bigg). \nonumber
\end{IEEEeqnarray}

As in the one-dimensional case, we now have a probability density with respect to $H_x$ and $H_y$, which we need to convert to one with respect to $x$ and $y$. For that we need to multiply $p$ with the Jacobian determinant
\begin{equation}
  \bigg| \frac{\partial(H_x,H_y)}{\partial(x,y)} \bigg|  =  |H_{xx}H_{yy}-H_{xy}^2|.
  \label{eq_Jacobian_2D}
\end{equation}
The probability distribution for the minima is thus given by
\begin{IEEEeqnarray}{rLlll}
  g(z)	& =	& \mc{3}{l}{ \frac1{n} \, p(H_x = 0) \, p(H_y = 0) } \nl
  	&	& \times \iiint	& \mc{2}{l}{ \rd H_{xx} \rd H_{yy} \rd H_{xy} \, p(H = z, H_{xx}, H_{yy}) } \nl
  	&	&		& \mc{2}{l}{ \times \: p(H_{xy}) \, |H_{xx}H_{yy}-H_{xy}^2| } \nl
  	& =	& \mc{2}{l}{ \frac1{n\pi K_2} \iiint }	& \rd H_{xx} \rd H_{yy} \rd H_{xy} \, p(z, H_{xx}, H_{yy}) \nl
  	&	&		&			& \times \: p(H_{xy}) \, |H_{xx}H_{yy}-H_{xy}^2|.
\end{IEEEeqnarray}
The integrals must be taken over the volume for which $H_{xx}H_{yy} - H_{xy}^2 > 0$ and $H_{xx}, H_{yy} > 0$, which forms the domain of the minima. These constraints and the integration can be simplified by making the following change of variables,
\begin{IEEEeqnarray}{rL}
  r\cos\theta				& = \tfrac12(H_{xx}-H_{yy}), \IEEEyessubnumber \\
  r\sin\theta				& = H_{xy}, \IEEEyessubnumber \\
  s					& = \tfrac12(H_{xx}+H_{yy}), \IEEEyessubnumber \\
  \rd H_{xx} \rd H_{yy} \rd H_{xy}	& = 2r \, \rd r \rd s \rd \theta.
\end{IEEEeqnarray}
In terms of these new variables, we have $H_{xx}H_{yy}-H_{xy}^2 = s^2-r^2$ and the constraints of the volume are given by $0 < r < s$. We get
\begin{equation}
  \begin{split}
    g(z) =\:	& \frac1{n\pi K_2} \int_0^{2\pi} \!\! \int_0^\infty \!\! \int_0^s \rd r \rd s \rd \theta \, \frac{4r(s^2-r^2)}{\pi^2K_4\sqrt{K_4-K_2^2}} \\
    		& \times \exp\bigg( \!-\! \frac{K_4z^2+4K_2sz+4s^2}{2(K_4-K_2^2)} - \frac{4r^2}{K_4} \bigg).
  \end{split}
\label{eq:integral}
\end{equation}
The density of the minima $n$ can again readily be obtained by integrating over $z$:
\begin{equation}
  \int_{-\infty}^\infty \! \rd z \, g(z)  =  1  \quad \Rightarrow \quad  n  =  \frac{K_4}{8\sqrt3 \pi K_2}.
  \label{eq_density_minima}
\end{equation}
Note that this result matches the one obtained in \cite{cite_Longuet2}.

\begin{figure}
  \centering
  \subfigure[]{\includegraphics{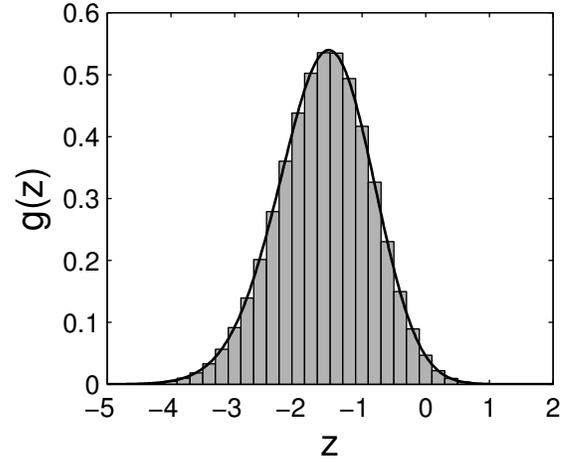}}
  \subfigure[]{\includegraphics{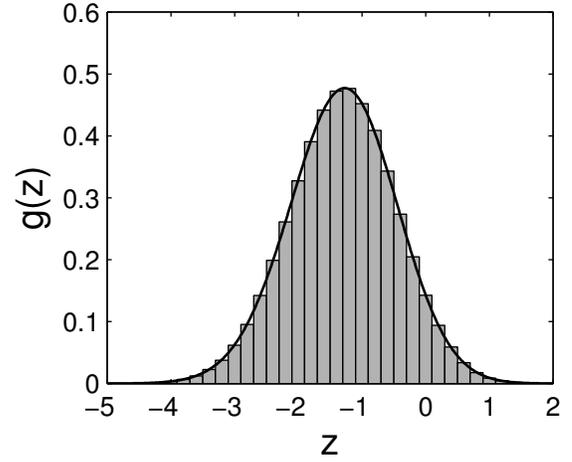}}
  \caption{Histograms of the values of $10^6$ minima obtained from simulations, together with the distribution given by eq.~\eqref{eq_distr}, for (a) a disk spectrum ($\lambda = \tfrac34$); (b) a Gaussian spectrum ($\lambda = \tfrac12$).}
  \label{fig_distr+hist}
\end{figure}

After evaluating the double integral (taking care to integrate over $r$ first), we obtain
\begin{widetext}
  \begin{equation}
    \begin{split}
      g(z) =\:	& \sqrt{\frac{3}{2\pi(3-2\lambda)}} \exp\bigg( \!-\! \frac3{2(3-2\lambda)} \, z^2\bigg) \erfc\Bigg( \sqrt{\frac{\lambda}{2(1-\lambda)(3-2\lambda)}} \, z \Bigg) \\
      		& - \sqrt{\frac{3}{2\pi}} \lambda (1-z^2) \exp\big( \!-\! \tfrac12 z^2 \big) \erfc\Bigg( \sqrt{\frac{\lambda}{2(1-\lambda)}} \, z \Bigg) \\
      		& - \frac1{\pi} \sqrt{3\lambda(1-\lambda)} \, z \exp\bigg( \!-\! \frac1{2(1-\lambda)} \, z^2 \bigg).
    \end{split}
    \label{eq_distr}
  \end{equation}
\end{widetext}
The two parameters $K_2$ and $K_4$ have been merged into one as before,
\begin{equation}
  \lambda  \equiv  \frac{K_2^2}{K_4}  \qquad  (0 \leq \lambda \leq 1).
\end{equation}
Again, when we set $K_0 = \la H^2 \ra \neq 1$, we get $\lambda = K_2^2/(K_0K_4)$.

Let us prove that $\lambda \leq 1$. After some rearranging, we see that this is equivalent to $K_0 K_4 - K_2^2 \geq 0$. We find
\begin{equation}
  K_0 K_4 - K_2^2  =  \iint \rd k \rd k' \, \Pi(k) \Pi(k') (k'^4 - k^2 k'^2).
\end{equation}
Note that we could just as well replace $k'^4$ with $k^4$ (because everything else is symmetric in $k$ and $k'$), and hence also with $\tfrac12 (k^4 + k'^4)$. If we do the latter, we can rewrite
\begin{equation}
  \tfrac12 (k^4 + k'^4) - k^2 k'^2  =  \tfrac12 (k - k')^2.
\end{equation}
We see that this is positive, together with $\Pi(k)$ and $\Pi(k')$, hence the integrand is positive and the integral too, which concludes the proof.

We have compared eq.~\eqref{eq_distr} with distributions obtained from computer-generated Gaussian fields -- details about these numerical simulations and how the minima were identified can be found in appendix~\ref{app_simulations}.
As can be seen in fig.~\ref{fig_distr+hist}, the agreement between eq.~\eqref{eq_distr} and the numeric results is excellent.

\begin{figure}
  \centering
  \includegraphics{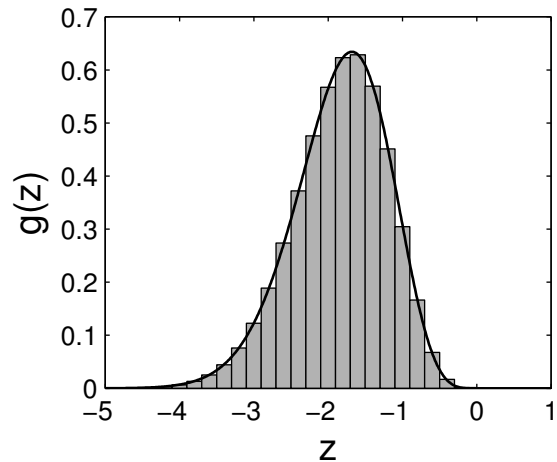}
  \caption{Histogram of the values of $10^6$ minima obtained from simulations, together with the distribution given by eq.~\eqref{eq_distr}, for a ring spectrum ($\lambda = 1$). No minima with a positive value of $H$ were found.}
  \label{fig_distr+hist_ring}
\end{figure}

Let us take a closer look at eq.~\eqref{eq_distr}. The two limits of $\lambda$ give results with interesting physical interpretations:
\begin{align}
  \lim_{\lambda \rightarrow 0} g(z)	& = \frac1{\sqrt{2\pi}}e^{-\frac12 z^2} - \sqrt{\lambda}\frac4{\sqrt3 \pi}ze^{-\frac12 z^2} + O(\lambda),	\label{eq_limit_zero} \\
  \lim_{\lambda \rightarrow 1} g(z)	& = (1-\sgn z)\sqrt{\frac3{2\pi}}\big(e^{-z^2}-1+z^2\big)e^{-\frac12 z^2}.					\label{eq_limit_one}
\end{align}
The case $\lambda = 0$ occurs when $K_4$ is unbounded (e.g.\ when $\Pi(k)$ scales as $k^{-6}$). We see that the distribution is then an elementary Gaussian. A rough intuitive explanation for this is as follows. The key feature of this limit is that the maxima and minima arise from very rapid oscillations that are superimposed on top of a slowly-varying field. In fact, if $K_4$ is extremely large, the waves with a short wavelength (large $|\vec{k}|$) have an amplitude that is small, but not negligible. They therefore create large fluctuations in the gradient of the field and hence a lot of extrema; a fact that can also be seen from eq.~\eqref{eq_density_minima}. 
Meanwhile, the height of the surface at any point (including the abundant minima) is dominated by the waves with a large amplitude, which have long wavelengths (small $|\vec{k}|$). The location of the minima and the height of the surface are thus \emph{independent}. Therefore, the distribution of the value of $H$ at a minimum is the same as for any other point: Gaussian. 

Now we consider $\lambda = 1$. From our proof that $\lambda \leq 1$, it is not hard to see that this can only occur when $\Pi(k) = \delta(k - k_0)$ for some constant $k_0$. This is called a ring spectrum, since the only occurring wave vectors are the ones with $|\vec{k}| = k_0$, which describes a circle in $\vec{k}$-space. Inspecting eq.~\eqref{eq_limit_one} we see that, due to the factor $(1-\sgn z)$, all minima have a negative value of $H$, as the simulations also show (see fig.~\ref{fig_distr+hist_ring}).
The explanation is that height fields with a ring spectrum necessarily satisfy $\nabla^2 H=-k_0^2 H$ -- therefore, if $H$ is positive, the mean curvature $H_{xx}+H_{yy}<0$, so the point cannot be a minimum. In other words, such Gaussian fields are random solutions to Helmholtz's equation -- they could represent the height field of a large membrane resonating at a certain frequency but with some randomness preventing a particular mode among the many at that frequency from stabilizing.

\begin{figure}
  \centering
  \includegraphics{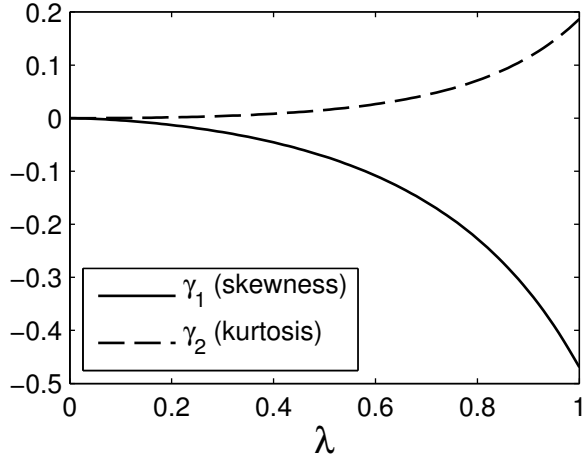}
  \caption{The skewness ($\gamma_1$) and kurtosis ($\gamma_2$) of the distribution (eq.~\eqref{eq_distr}) as a function of $\lambda$ (see eqs.~\eqref{eq_skewness} and \eqref{eq_kurtosis}).}
  \label{fig_skewness+kurtosis}
\end{figure}

While eq.~\eqref{eq_distr} appears quite complex, some of its parameters have more transparent forms. The expectation value $\mu$ and standard deviation $\sigma$ for example are
\begin{align}
  \mu		& = -4\sqrt{\frac2{3\pi}\lambda},			\label{eq_mu} \\
  \sigma	& = \sqrt{1 - \frac{32-(6\sqrt3-2)\pi}{3\pi}\lambda}.	\label{eq_sigma}
\end{align}

When looking at fig.~\ref{fig_distr+hist}, it appears that the distribution is itself almost Gaussian. This can be captured in the skewness $\gamma_1$ and kurtosis $\gamma_2$,
\begin{align}
  \gamma_1	& \equiv  \frac{\mu_3}{\sigma^3}  =  -\frac{ 4\sqrt2 \big( 64 - (18\sqrt3-11)\pi \big) }{ \big( 3\pi\lambda^{-1} - (32 - (6\sqrt3-2)\pi) \big) ^{3/2} }	\label{eq_skewness} \nl
  		& = -\frac{3.46}{(9.42\lambda^{-1}-5.63)^{3/2}}, \\
  \gamma_2	& \equiv  \frac{\mu_4}{\sigma^4}-3 \nl
  		& = \frac{ 4 \big( \!-\! 1536 + 32(18\sqrt3-11)\pi + 9(2\sqrt3-9)\pi^2 \big) }{ \big( 3\pi\lambda^{-1} - (32 - (6\sqrt3-2)\pi) \big) ^2 }	\label{eq_kurtosis} \nl
  		& = \frac{2.68}{(9.42\lambda^{-1}-5.63)^2}.
\end{align}
Here $\mu_n$ is the $n$-th moment about the mean: $\mu_n \equiv \la (\xi - \la\xi\ra)^n \ra$. The skewness is a measure of the symmetry of a distribution around the mean, while the kurtosis gives an indication of its ``peakiness''. For a Gaussian distribution, both the skewness and the kurtosis are zero. They can therefore be considered as a measure of the Gaussianity of a distribution; note however that a distribution is not necessarily Gaussian if both parameters are zero. The two parameters are shown in fig.~\ref{fig_skewness+kurtosis}. Naturally, they both go to zero for $\lambda \rightarrow 0$.

		\section{Maxima and Minima Imbalance}
	\label{sec_4}

\begin{figure}
  \centering
  \subfigure[]{\includegraphics{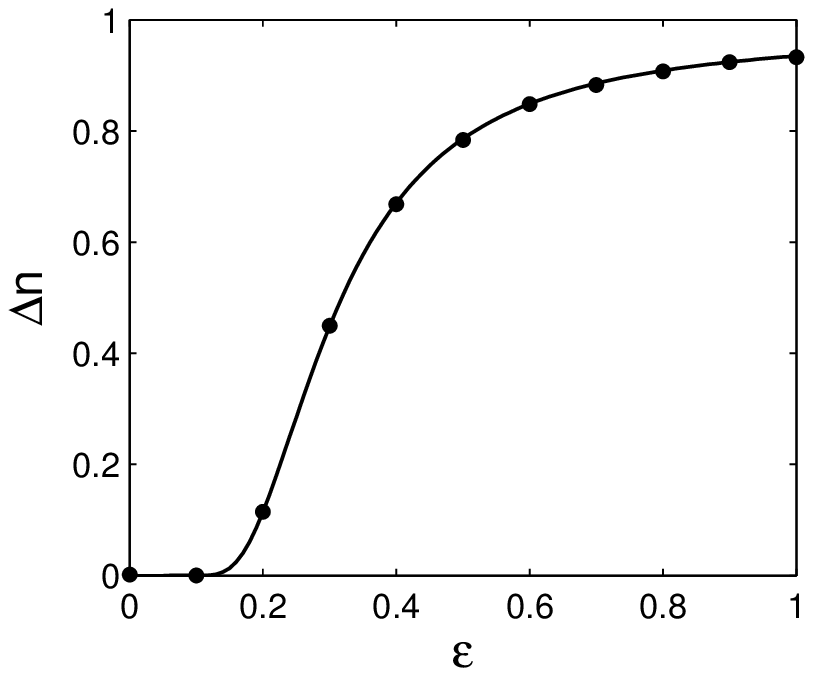}}
  \subfigure[]{\includegraphics{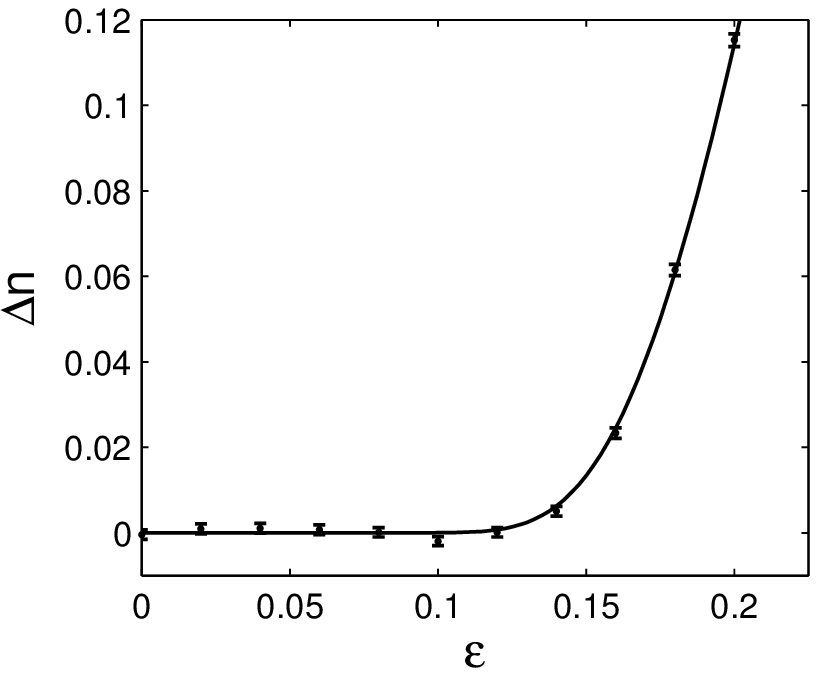}}
  \caption{$\Delta n$ for $h = H + \eps H^2$ as a function of $\eps$, where $H$ has a disk spectrum ($\lambda = \frac34$). The data points stem from simulations, the solid curve is eq.~\eqref{eq_maxmin}. The two graphs are for different ranges of $\eps$.}
  \label{fig_gsn+sq_disk_maxmin+theory}
\end{figure}

\begin{figure}
  \centering
  \subfigure[]{\includegraphics{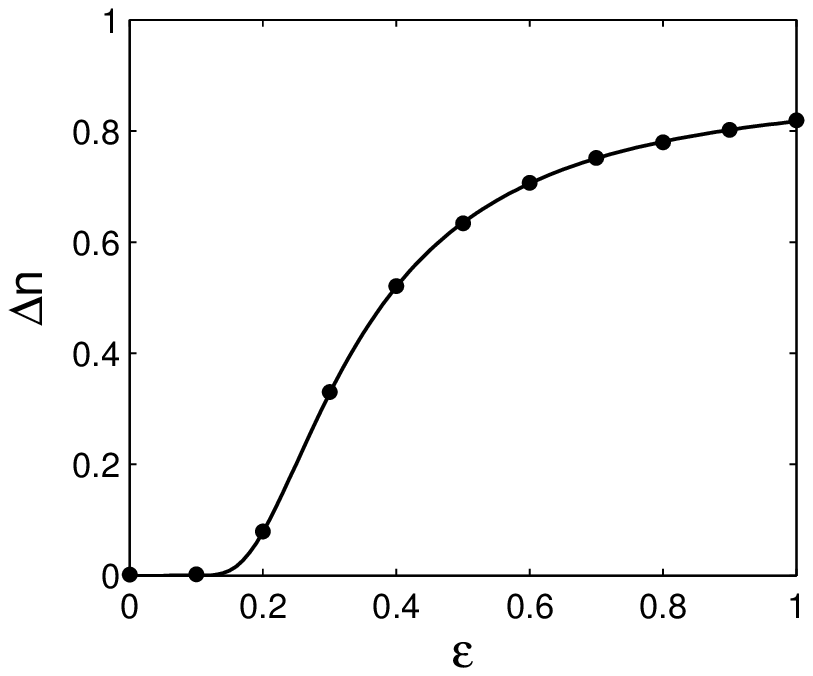}}
  \subfigure[]{\includegraphics{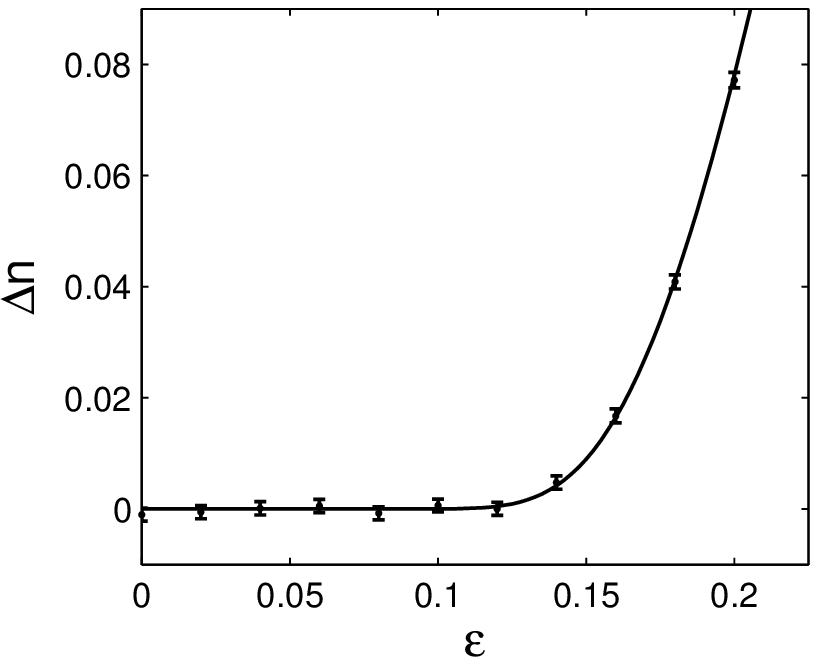}}
  \caption{$\Delta n$ for $h = H + \eps H^2$ as a function of $\eps$, where $H$ has a Gaussian spectrum ($\lambda = \tfrac12$). The data points stem from simulations, the solid curve is eq.~\eqref{eq_maxmin}. The two graphs are for different ranges of $\eps$.}
  \label{fig_gsn+sq_gauss_maxmin+theory}
\end{figure}

Now that we have obtained $g(z)$, we can calculate the relative imbalance between the densities of maxima and minima of $h = F_{NL}(H)$, in accordance with eq.~\eqref{eq_maxmin1}:
\begin{equation}
  \Delta n  \equiv  \frac{n_{\mathrm{max}}-n_{\mathrm{min}}}{n_{\mathrm{max}}+n_{\mathrm{min}}}  =  \int_{z:F_{NL}'(z)<0} \! \rd z \, \big( g(z) - g(-z) \big).
  \label{eq_maxmin}
\end{equation}
The most basic example of a perturbed Gaussian for which we may expect $\Delta n \neq 0$ is $h = H + \eps H^2$. In this case, the domain of integration is $[-\infty, -\frac1{2\eps}]$. We have compared eq.~\eqref{eq_maxmin} with results from computer-generated fields, for two different spectra: in fig.~\ref{fig_gsn+sq_disk_maxmin+theory} a so-called disk spectrum was used:
\begin{equation}
  A(k)^2 \sim \theta(k - k_0)  \qquad  K_{2n} = \frac{k_0^{2n}}{n+1}  \qquad  \lambda = \frac34
\end{equation}
Fig.~\ref{fig_gsn+sq_gauss_maxmin+theory} features results for a Gaussian spectrum:
\begin{equation}
  A(k)^2 \sim \exp(-k^2 / 2k_0^2)  \qquad  K_{2n} = 2^n n! k_0^{2n}  \qquad  \lambda = \frac12
\end{equation}

In both cases, we see an excellent agreement between the results from the simulations and our theoretical formula.

In both figures, we see that $\Delta n$ increases dramatically starting $\eps \sim 0.15$. This can be explained intuitively as follows: the balance in densities of maxima and minima is disturbed by extrema located below $H = -\frac1{2\eps}$. Since $H$ is a standard Gaussian, such low values (i.e.\ large negative values) of $H$ are exponentially rare. It is only when $-\frac1{2\eps}$ is in the order of $-1$ that a significant $\Delta n$ can be expected. To get a rough estimate for the number of these extrema, we can just look at the density of points with $H=-\frac{1}{2\eps}$ (ignoring the requirement that they be minima does not change the exponential dependence). This is $e^{-1/(8\eps^2)}$. A more careful approximation (see appendix~\ref{app_asymptotes}) gives $\Delta n\sim\sqrt{\frac{3}{2\pi}}\frac{\lambda}{\eps}e^{-\frac{1}{8\eps^2}}$.

This argument also applies to the generic case $h = H + \eps f_{NL}(H)$, where $f_{NL}$ designates a perturbation and $\eps$ is a parameter controlling the size of the perturbation. Now $\eps f_{NL}'(H)$ needs to be in the order of $1$ for $\Delta n$ to be significantly nonzero. Thus measuring the imbalance between maxima and minima does not give a very sensitive test of the type of non-Gaussianity that we have considered here, in the limit of small $\eps$. However, eq.~\eqref{eq_maxmin} is a nonperturbative result that also holds for large $\eps$.

		\section{Conclusions}
	\label{sec_5}

For a random field given by $h(\vec{r}) = F_{NL}(H(\vec{r}))$, where $H$ is a Gaussian field and $F_{NL}$ any (nonlinear) function, we find that the densities of maxima and minima of $h$ may differ. We have shown what the imbalance is as a function of the transformation $F_{NL}$ and the power spectrum of $H$. Our result is exact, and does not rely on perturbation theory, a nice feature since $F_{NL}$ does not have to be small for our result to apply. This is confirmed by our simulations. On the other hand, the imbalance between maxima and minima is exponentially small when $\eps$ is small. Directly measuring the skewness at a given point, for example is much more sensitive to $\eps$ when $\eps \ll 1$.

The simple reason is that, when $H$ is of order one, $h$ is a monotonic function of $H$, and hence it has the same numbers of maxima and minima. Only very large fluctuations in $H$ can lead to an imbalance. Other types of non-Gaussian fields are more likely to have appreciable imbalances between maxima and minima. For example, the nonlinear evolution of a field, such as the height of a surface on which particles are accumulating could give rise to an imbalance between maxima and minima. The diffusion of the particles for instance might preferentially smooth out maxima.



\appendix

		\section{Properties of Gaussian fields}
	\label{app_gsn_prop}

Let us start by calculating the mean and standard deviation of a Gaussian field $H$, the equivalents of $\mu$ and $\sigma$ of a Gaussian variable. This involves expectation values, which are obtained by integrating over all possible values of all random variables, which in this case, are the uniformly distributed phases
\begin{equation}
  \la \ldots \ra  \equiv  \bigg( \prod_{\vec{k}} \int \! \frac{\rd \phi_{\vec{k}}}{2\pi} \bigg) \ldots.
\end{equation}
The mean is then simply
\begin{align}
  \la H(\vec{r}) \ra	& = \Big\la H_0 + \sum_{\vec{k}} A(k) \cos(\vec{k} \cdot \vec{r} + \phi_{\vec{k}}) \Big\ra \nl
  			& = H_0 + \sum_{\vec{k}} A(k) \big\la \! \cos(\vec{k} \cdot \vec{r} + \phi_{\vec{k}}) \big\ra \nl
  			& = H_0 + \sum_{\vec{k}} A(k) \int \! \frac{\rd \phi_{\vec{k}}}{2\pi} \cos(\vec{k} \cdot \vec{r} + \phi_{\vec{k}}) \nl
  			& = H_0.
\end{align}

For the variance (standard deviation squared) we find
\begin{align}
  & \la (H - \la H \ra)^2 \ra \nl
  & \qquad = \Big\la \Big( \sum_{\vec{k}} A(k) \cos(\vec{k}\cdot\vec{r} + \phi_{\vec{k}}) \Big)^2 \Big\ra \nl
  & \qquad = \sum_{\vec{k}\vec{k'}} A(k)A(k') \big\la \! \cos(\vec{k}\cdot\vec{r} + \phi_{\vec{k}}) \cos(\vec{k'}\cdot\vec{r} + \phi_{\vec{k'}}) \big\ra.
\end{align}
Since the phases are uncorrelated, for $\vec{k} \neq \vec{k'}$ we find
\begin{align}
  & \big\la \! \cos(\vec{k}\cdot\vec{r} + \phi_{\vec{k}}) \cos(\vec{k'}\cdot\vec{r} + \phi_{\vec{k'}}) \big\ra \nl
  & \qquad = \big\la \! \cos(\vec{k}\cdot\vec{r} + \phi_{\vec{k}}) \big\ra \big\la \! \cos(\vec{k'}\cdot\vec{r} + \phi_{\vec{k'}}) \big\ra  =  0.
\end{align}
Hence the term in the double sum can only be nonzero for $\vec{k} = \vec{k'}$. As a result we get
\begin{align}
  \la (H - \la H \ra)^2 \ra	& = \sum_{\vec{k}} A(k)^2 \big\la \! \cos^2(\vec{k}\cdot\vec{r} + \phi_{\vec{k}}) \big\ra \nl
  				& = \sum_{\vec{k}} \tfrac12 A(k)^2.
\end{align}

For simplicity, we will set $\la H \ra = 0$ and $\la H^2 \ra = 1$, which translates to $H_0 = 0$ and $\sum_{\vec{k}} \tfrac12 A(k)^2 = 1$.

While the vectors $\vec{k}$ in eq.~\eqref{eq_gsn_field} form a discrete set, usually they are sufficiently finely spaced so that we can treat the amplitude spectrum $A(k)$ as a continuous function defined over the positive reals. If we take our normalization condition, and replace the sum with an integral, we get
\begin{align}
  1	& = \sum_{\vec{k}} \tfrac12 A(k)^2  =  \int \! \rd \vec{k} \, \tfrac12 a(k)^2  =  \int_0^{2\pi} \! \int_0^\infty \! k \, \rd k \rd \theta \, \tfrac12 a(k)^2 \nl
  	& = \int_0^\infty \! \rd k \, \pi k a(k)^2  =  \int_0^\infty \! \rd k \, \Pi(k).
\end{align}
Here $a(k)$ indicates the continuous spectrum equivalent to the discrete amplitudes $A(k)$. The newly introduced function $\Pi(k) \equiv \pi k a(k)^2$ is the \emph{power spectrum} of $H$.

Some properties of a Gaussian field depend on the amplitude spectrum. In many cases this dependence can be expressed in terms of the \emph{moments} of the spectrum
\begin{equation}
  K_n  =  \sum_{\vec{k}} \tfrac12 A(k)^2 k^n  =  \int_0^\infty \! \rd k \, \Pi(k) k^n.
  \label{eq_moments}
\end{equation}
The normalization condition can be translated as $K_0 = 1$.

	\subsection{Two-point correlation function}

Correlation functions are often used to probe the Gaussianity of a given random field. This is because for Gaussian fields, they obey certain relations, as reviewed below. We will first calculate the two-point correlation function.

The two-point correlation function $C(\vec{r_1}, \vec{r_2})$ of a field $\psi$ is defined as
\begin{equation}
  C(\vec{r_1}, \vec{r_2})  =  \la \psi(\vec{r_1}) \psi(\vec{r_2}) \ra.
\end{equation}
When $\psi$ is homogeneous and isotropic, $C$ depends only on the distance between $\vec{r_1}$ and $\vec{r_2}$
\begin{equation}
  C(R)  =  \la \psi(\vec{r}) \psi(\vec{r} + \vec{R}) \ra,
\end{equation}
where $\vec{r}$ is any position and $\vec{R}$ is any vector of length $R$. For a Gaussian field $H$ we find (if we set $\vec{r} = 0$ for convenience, which we are free to do)
\begin{align}
  & C(R) = \la H(0) H(\vec{R}) \ra \nl
  & \quad = \bigg\la \Big( \sum_{\vec{k}} A(k) \cos(\phi_{\vec{k}}) \Big) \Big( \sum_{\vec{k}} A(k) \cos(\vec{k} \cdot \vec{R} + \phi_{\vec{k}}) \Big) \bigg\ra \nl
  & \quad = \sum_{\vec{k}\vec{k'}} A(k) A(k') \big\la \! \cos(\phi_{\vec{k}}) \cos(\vec{k'} \cdot \vec{R} + \phi_{\vec{k'}}) \big\ra.
\end{align}
Since the phases $\phi_{\vec{k}}$ are uncorrelated, the correlation is automatically zero when $\vec{k} \neq \vec{k'}$, hence
\begin{align}
  C(R)	& = \sum_{\vec{k}} A(k)^2 \big\la \! \cos(\phi_{\vec{k}}) \cos(\vec{k} \cdot \vec{R} + \phi_{\vec{k}}) \big\ra \nl
  	& = \sum_{\vec{k}} A(k)^2 \big\la \tfrac12 \cos(\vec{k} \cdot \vec{R} + 2\phi_{\vec{k}}) + \tfrac12 \cos(\vec{k} \cdot \vec{R}) \big\ra.
\end{align}
Since $\phi_{\vec{k}}$ is uniformly distributed, the expectation value of the first cosine is zero, and we are left with
\begin{equation}
  C(R)  =  \sum_{\vec{k}} \tfrac12 A(k)^2 \cos(\vec{k} \cdot \vec{R})  =  \int \! \rd \vec{k} \, \tfrac12 a(k)^2 \cos(\vec{k} \cdot \vec{R}).
\end{equation}
We thus find that the two-point correlation function of a Gaussian field is the Fourier transform of its (two-dimensional) power spectrum. Therefore, in essence, the correlation function is as much a complete description of a Gaussian field as the power spectrum is. Also, by determining the correlation function and taking the inverse Fourier transform, one obtains the spectrum.

The moments, defined before in terms of the power spectrum, can be related to the derivatives of the correlation function at $R = 0$. Because of symmetry, we must have $C(R) = C(-R)$. Hence $C(R)$ is an even function and all its odd derivatives at zero vanish. To obtain the even derivatives, we must first eliminate the vector $\vec{R}$ in the equation above. We are free to choose its direction, so let us take $\vec{R} = R\hat{x}$. We then get
\begin{align}
  C^{(2n)}(R)	& = \bigg(\frac{\rd}{\rd R}\bigg)^{2n} \int \! \rd \vec{k} \, \tfrac12 a(k)^2 \cos(k_x R) \nl
  		& = \int \! \rd \vec{k} \, \tfrac12 a(k)^2 (-1)^n k_x^{2n} \cos(k_x R), \\
  C^{(2n)}(0)	& = (-1)^n \int \! \rd \vec{k} \, \tfrac12 a(k)^2 k_x^{2n} \nl
  		& = (-1)^n \int_0^{2\pi} \! \int_0^\infty \! k \, \rd k \rd \theta \, \tfrac12 a(k)^2 k^{2n} \cos^{2n} \theta \nl
  		& = (-1)^n K_{2n} \frac1{2\pi} \int \! \rd \theta \, \cos^{2n} \theta \nl
  		& = (-1)^n \frac{(2n-1)!!}{2^n n!} K_{2n}.
\end{align}
We thus find a one-to-one relation between the moments and the derivatives of the correlation function. The derivative of the correlation function $C^{(2n)}(0)$ is related to roughness in the field itself. In fact, $C^{(2n)}(0)$ is equal to $(-1)^n \la [H^{(n)}(x)]^2\ra$, the fluctuations of the $n$-th derivative (apart from a sign).

	\subsection{Higher order correlation functions}

In general, the $n$-point correlation function is defined as the expectation value $\la \psi(\vec{r_1}) \psi(\vec{r_2}) \ldots \psi(\vec{r_n}) \ra$, as a function of $\vec{r_1}$ through $\vec{r_n}$. For a Gaussian field, this correlation function can be expressed in terms of two-point correlation functions, analogous to Wick's theorem. As a result, a non-Gaussian field can be recognized by checking whether this relation holds. In practice, this test can be applied to a single Gaussian field if it is homogeneous. In that case, the correlation function depends only on separations of the points $\vec{r_2} - \vec{r_1}$ through $\vec{r_n} - \vec{r_1}$. From a given homogeneous field $\psi$, one obtains (a good approximation of) this correlation function by averaging over all (or a lot of) configurations with fixed spacings but translated to different $\vec{r_1}$'s.

The simplest case of the relationship is
\begin{equation}
  \begin{split}
    \la H_1 H_2 H_3 H_4 \ra =\:	& \la H_1 H_2 \ra \la H_3 H_4 \ra + \la H_1 H_3 \ra \la H_2 H_4 \ra \\
    						& + \la H_1 H_4 \ra \la H_2 H_3 \ra,
  \end{split}
  \label{eq_four_pt_corr}
\end{equation}
where we introduced the notation $H_i \equiv H(\vec{r_i})$ for shortness. In general, correlations between an even number of variables with $n>2$ can be reduced to the two-point correlations, while correlations between an odd number of variables always vanish. These properties follow from the definition of the Gaussian field: the $n$ variables $H(r_i)$ are described by a correlated Gaussian distribution, and hence their correlation functions can be calculated explicitly from Gaussian integrals.


We shall now show how this characteristic relation comes about for our Fourier superposition. When we calculate the four-point correlation in the same way as we did for the two-point correlation, we bring the brackets inside the (quadruple) sum, which gives us the term
\begin{equation}
  \begin{split}
    \big\la \!	& \cos(\vec{k_1} \cdot \vec{r_1} + \phi_{\vec{k}_1}) \cos(\vec{k_2} \cdot \vec{r_2} + \phi_{\vec{k}_2}) \\
    		& \times \cos(\vec{k_3} \cdot \vec{r_3} + \phi_{\vec{k}_3}) \cos(\vec{k_4} \cdot \vec{r_4} + \phi_{\vec{k}_4}) \big\ra,
  \end{split}
\end{equation}
which is summed for all combinations of $\vec{k_1}$ through $\vec{k_4}$. The first thing to note, is that whenever e.g.\ $\vec{k_1}$ is not equal to any of the other $\vec{k_i}$, the correlation is automatically zero; this is because $\cos(\vec{k_1} \cdot \vec{r_1} + \phi_{\vec{k}_1})$ is then independent of all other factors, can therefore be separated, and gives zero. Hence the correlation can only be nonzero if each $\vec{k_i}$ is equal to (at least) one other $\vec{k_i}$. We can distinguish the cases $\vec{k_1} = \vec{k_2}, \vec{k_3} = \vec{k_4}$ and $\vec{k_1} = \vec{k_3}, \vec{k_2} = \vec{k_4}$ and $\vec{k_1} = \vec{k_4}, \vec{k_2} = \vec{k_3}$.
Let us focus on the first case; the sum of all these correlations gives
\begin{equation}
  \begin{split}
    \sum_{\vec{k_1}, \vec{k_3}}	& A(k_1)^2 A(k_3)^2 \big\la \! \cos(\vec{k_1} \cdot \vec{r_1} + \phi_{\vec{k}_1}) \cos(\vec{k_1} \cdot \vec{r_2} + \phi_{\vec{k}_1}) \\
    				& \qquad \times \cos(\vec{k_3} \cdot \vec{r_3} + \phi_{\vec{k}_3}) \cos(\vec{k_3} \cdot \vec{r_4} + \phi_{\vec{k}_3}) \big\ra.
  \end{split}
\end{equation}
This can be split into
\begin{equation}
  \begin{split}
    &        \sum_{\vec{k_1}} A(k_1)^2 \big\la \! \cos(\vec{k_1} \cdot \vec{r_1} + \phi_{\vec{k}_1}) \cos(\vec{k_1} \cdot \vec{r_2} + \phi_{\vec{k}_1}) \big\ra \\
    & \times \sum_{\vec{k_3}} A(k_3)^2 \big\la \! \cos(\vec{k_3} \cdot \vec{r_3} + \phi_{\vec{k}_3}) \cos(\vec{k_3} \cdot \vec{r_4} + \phi_{\vec{k}_3}) \big\ra \\
    & \qquad = \la H(\vec{r_1}) H(\vec{r_2}) \ra \la H(\vec{r_3}) H(\vec{r_4}) \ra.
  \end{split}
\end{equation}
Applying the same to the other cases and adding them together precisely gives eq.~\eqref{eq_four_pt_corr}.

One may note that the case $\vec{k_1} = \vec{k_2} = \vec{k_3} = \vec{k_4}$ has not been treated correctly. However, since $\vec{k_i}$ can take on an infinite number of values, and this case only provides one degree of freedom instead of the two we had for the other cases, these correlations only have an infinitesimal contribution.

From this example, it is not hard to see that in general, an $n$-point correlation function can be factorized, that is, written as the sum of products of two-point correlations, where the sum features all possible ways in which the $n$ variables can be paired up.

		\section{Computer simulations}
	\label{app_simulations}

\begin{figure}
  \centering
  \includegraphics{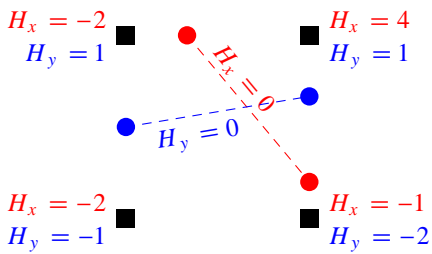}
  \caption{Identifying a critical point. The four black squares are grid points at which $H_x$ and $H_y$ are known. At the red (blue) dots, $H_x = 0$ ($H_y = 0$) under the assumption that $H_x$ ($H_y$) is linear between the two grid points. The two contour lines $H_x = 0$ and $H_y = 0$ then intersect inside the square, indicating the existence of a critical point.}
  \label{fig_crit_ident}
\end{figure}

In order to verify our theoretical results, we use a large number of computer-generated realizations of the Gaussian field $H$ (typically a few thousand) -- each with the same spectrum $A(k)$ but random phases $\phi_{\vec{k}}$. We then apply the desired transformation $F_{NL}$ and extract the desired statistics.

The fields are defined on a square with periodic boundary conditions, i.e.\ $H(x,y) = H(x+L,y) = H(x,y+L)$, in order to reduce finite size effects. This is accomplished by only adding together waves (as in eq.~\eqref{eq_gsn_field}) with wave vectors $\vec{k}$ of which the $x$ and $y$-components are multiples of $2\pi/L$.

The summation in eq.~\eqref{eq_gsn_field} is restricted to wave vectors with a magnitude below a certain threshold $k_{max}$. In order to minimize the potential effects of this cutoff, we choose spectra for which $A(k)$ decays very quickly or is zero for large $k$, such as the disk spectrum
\begin{equation}
  A(k)^2 \propto \theta(k - k_0),
\end{equation}
and the Gaussian spectrum
\begin{equation}
  A(k)^2 \propto \exp(-k^2 / 2k_0^2).
\end{equation}

Finally, $L$ is chosen in relation to $k_{max}$ such that (1) the sum in eq.~\eqref{eq_gsn_field} features at least a few hundred waves (recall that $L$ influences this number via the periodic boundary conditions) and that (2) $L$ is at least a few times $2\pi/\sqrt{K_2}$, which is a measure of the typical wavelength of the spectrum.
An example of a Gaussian field generated in this way is shown in fig.~\ref{fig_surface_disk}.

The resulting formula for $H$ is then evaluated at the grid points only. The distance between neighboring grid points is taken to be much smaller (by a factor of $50$ roughly) than the typical wavelength. Along with $H$ itself, we also calculate its first and second derivatives at these grid points.

We use a very efficient approach to identifying critical points and their type (maximum, minimum or saddle point). Every square of four neighboring grid points is considered. If $H_x$ or $H_y$ has the same sign at all four points, we infer that it is not zero anywhere inside the square, which leads to the conclusion that the square does not contain a critical point.
Otherwise, there would necessarily be at least two pairs of neighboring grid points with a different sign of $H_x$. For each pair, it is assumed that $H_x$ changes linearly between the two points, which allows to pinpoint two points along the edges of the square where $H_x = 0$. The contour line $H_x = 0$ is then assumed to be a straight line between these two points. The same recipe is applied to $H_y$, after which it is determined whether the two contour lines crossed. The intersection (if present) is then a critical point. This idea is illustrated in fig.~\ref{fig_crit_ident}.

It is also possible for all four neighboring points to have opposite signs of $H_x$ (or $H_y$). This results in four points along the border of the square with $H_x = 0$, but without any information about which two pairs should be connected by a contour line. In combination with the two points with $H_y = 0$ it can however be established what the parity of the number of intersections (i.e.\ critical points) is. We then simply assume this number to be 0 or 1. This case is sufficiently rare (provided the grid is small enough) to not have a noticeable effect on the results.

Once established whether the square under consideration contains a critical point, the type is determined by averaging the values of $H_{xx}$, $H_{yy}$ and $H_{xy}$ at the four grid points and evaluating the signs of $H_{xx}H_{yy} - H_{xy}^2$ and $H_{xx} + H_{yy}$.

Although this method clearly does not always correctly determine the existence of a critical point or its type, it is not biased toward one outcome. Therefore, the mistakes that are made will get averaged out when statistics are taken over a large number of critical points. With regard to getting good statistics, the speed of this method is a big advantage.

This method -- along with the proper values of $k_{max}$, $L$ and the grid size -- was thoroughly tested on Gaussian fields (for which the statistical outcomes are known from theory) to verify its validity, before applying it to the non-Gaussian fields under investigation.

		\section{Asymptotes for a very small non-Gaussianity}
	\label{app_asymptotes}

In the limit where $z$ is very large and negative, $g(z)$ can be evaluated asymptotically. One can use the exact expression for $g(z)$, but returning to the original integral eq.~\eqref{eq:integral} gives more insight (and makes the calculations shorter). The exponential weight in the integral is peaked at $s=\frac{1}{2}(H_{xx}+H_{yy})=-\frac{K_2z}{2}$. Therefore $s$ almost certainly becomes large and positive when $-z$ is large, since the width of the distribution remains fixed. This allows us to extend the range of integration to all $s$'s and to all $r\geq 0$, since the additional parts of the range have a very small weight. Then the integral can be worked out exactly, giving $g(z)\approx\sqrt{\frac{6}{\pi}}\lambda z^2 e^{-\frac{z^2}{2}}$ apart from small corrections, when $-z$ is large and positive. We next substitute back in eq.~\eqref{eq_maxmin1}. Noting that $P$ dominates over $Q$ and using integration by parts to evaluate the integral asymptotically ($\int_A^\infty z^2e^{-\frac{z^2}{2}}dx\approx Ae^{-\frac{A^2}{2}}$ when $A\rightarrow\infty$) gives $\Delta n=\sqrt{\frac{3}{2\pi}}\frac{\lambda}{\eps}e^{-\frac{1}{8\eps^2}}$.



\bibliography{nongauss}

\end{document}